\documentclass[12pt,preprint]{aastex}

\newcommand\mdivi{\multicolumn{2}{c|}}
\newcommand\multi{\multicolumn{2}{c}}
\newcommand\divi{\multicolumn{1}{c|}}

\newcommand\up{\raisebox{4.7ex}[0pt]}
\newcommand\upa{\raisebox{1.7ex}[0pt]}
\newcommand\upb{\raisebox{2.7ex}[0pt]}
\newcommand\upc{\raisebox{0.5ex}[0pt]}
\newcommand\upd{\raisebox{-0.5ex}[0pt]}

\usepackage{amsmath}
\usepackage{graphicx}
\usepackage{graphics}
\usepackage{epsf}

\begin{document}

\title{Hardness Ratio Estimation in Low Counting X-ray Photometry}

\author{Y. K. Jin\altaffilmark{1}, S. N. Zhang\altaffilmark{2,3}, J. F. Wu\altaffilmark{2} }
\affil{\footnotesize $^{1}$Department of Engineering Physics and Center for
Astrophysics, Tsinghua University, Beijing 100084, P. R. China} \affil{\footnotesize
$^{2}$Department of Physics and Center for Astrophysics, Tsinghua University, Beijing
100084, P. R. China} \affil{\footnotesize $^{3}$Key Laboratory of Particle
Astrophysics, Institute of High Energy Physics, Chinese Academy of Sciences, Beijing
100049, P. R. China}

\begin{abstract}
Hardness ratios are commonly used in X-ray photometry to indicate
spectral properties roughly. It is usually defined as the ratio of
counts in two different wavebands. This definition, however, is
problematic when the counts are very limited. Here we instead
define hardness ratio using the $\lambda$ parameter of Poisson
processes, and develop an estimation method via Bayesian
statistics. Our Monte Carlo simulations show the validity of our
method. Based on this new definition, we can estimate the hydrogen
column density for the photoelectric absorption of X-ray spectra
in the case of low counting statistics.
\end{abstract}


\bigskip

\section{Introduction}

In recent years, high angular resolution X-ray telescopes make it
possible to detect X-ray sources with only a few counts. This is
very different from the optical photometry. Because of these low
counts, the Poisson processes in corresponding wavebands cannot be
approximated to Gaussian distribution. Therefore the statistics
will be very different in some estimations and calculations than
used before. In recent years Bayesian method has gained many
applications (e.g. van Dyk et al. 2001 and references therein)
since it has more advantages in low count cases than traditional
statistics.

Hardness ratios are widely used in high energy astrophysics since
faint sources with only limited counts cannot give any satisfying
spectral modeling. In X-ray detection, hardness ratios are
normally used to show spectral properties roughly (e.g. Tennant et
al., 2001; Sivakoff, Sarazin \& Carlin 2004). Hardness ratios are
usually defined as the ratio of counts in different wavebands
($HR=b/a$) or the ratio of the difference and sum of counts in two
wavebands ($HR=(b-a)/(b+a)$), $a$ and $b$ are counts in two
wavebands $A$ and $B$. On the other hand, for the spectra of X-ray
sources, the photoelectric absorption, quantified with the
hydrogen column density $N_H$, cannot be neglected. Hydrogen
column density contains many kinds of important information, such
as the radial distance and the interstellar circumstance of the
sources. For low count sources, hydrogen column density is hard to
know since no reliable spectral fitting can be made. However
interstellar absorption is energy dependent. Consequently the
information of hydrogen column density can be drawn from the
hardness ratios. In this paper, we first give a new definition of
hardness ratio and its estimation method, and then we discuss the
procedure to estimate the hydrogen column density accordingly.

\bigskip
\section{The distribution of $\lambda$ parameter under certain counts}

We begin our discussion with the following problem: Suppose that
one experiment obtained two counts from two different Poisson
distributions, we need to: (1) Estimate the ratio of the
expectation values of the two Poisson distributions, and (2)
Construct the confidence interval of the ratio.

The expectation values of the Poisson processes are just the
$\lambda$ parameters of the Poisson distribution
$P(n|\lambda)=\frac{\lambda^n}{n!}e^{-\lambda}$. Therefore the
above problem may be formulated as follows: Suppose $a$ and $b$
are two counts corresponding to two different Poisson processes
$A$ and $B$ with their parameters as $\lambda_A$ and $\lambda_B$
respectively, $\lambda_A/\lambda_B$ and its confidence interval
needs to be estimated. To solve this problem we first need to
derive the distribution of $\lambda$ parameter under certain
counts, i.e., the conditional distribution of $\lambda_A$ and
$\lambda_B$, as follows.

\begin{eqnarray}
p(\lambda_A=x|n_A=a)&=&\frac{P(n_A=a|\lambda_A=x)p(\lambda_A=x)}{\int_0^{\infty}P(n_A=a|\lambda_A=t)p(\lambda_A=t){\rm
d}t}.
\end{eqnarray}
First we assume, as a pragmatic convention, a uniform prior for
the $\lambda$ parameter.
\begin{eqnarray}p_u(\lambda_A=x|n_A=a)&=&\frac{P(n_A=a|\lambda_A=x)}{\int_0^{\infty}P(n_A=a|\lambda_A=t){\rm d}t}\nonumber\\
&=&\frac{x^ae^{-x}}{a!}.
\end{eqnarray}
Similarly,
\begin{eqnarray}p_u(\lambda_B=y|n_B=b)&=&\frac{y^be^{-y}}{b!}.
\end{eqnarray}

This continuous distribution is Gamma distribution, as shown in
Fig. 1.

\begin{figure}[tb]
\centerline{\epsfxsize 80mm \plotone{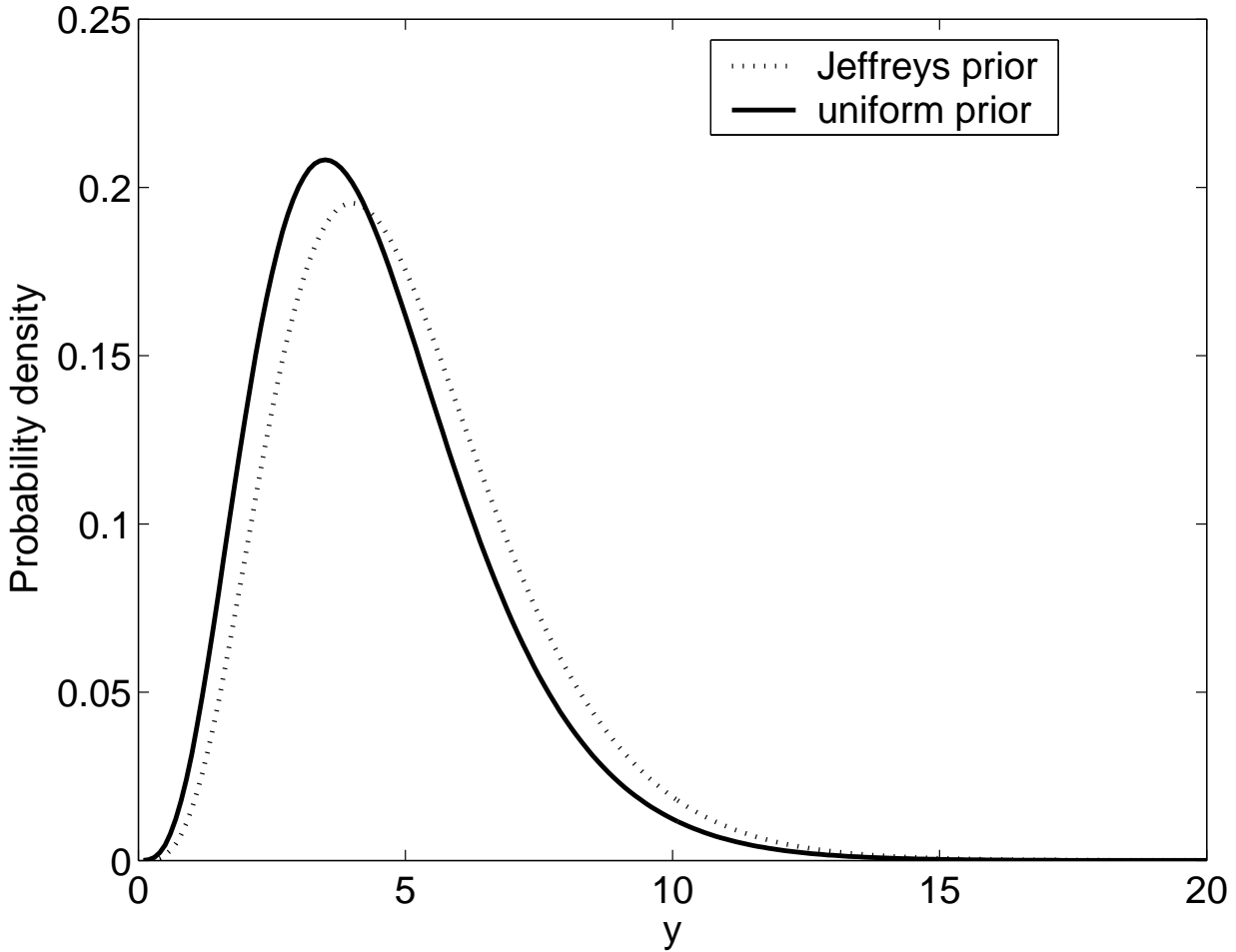}} \caption[]{{The
probability density function of $\lambda_B$ when $b=4$ under the
two different prior assumption.}} \label{f1}
\end{figure}

In addition, we use Jeffreys prior ($p(\lambda) = 1/\sqrt{\lambda}$), which may be more
advantageous over the uniform prior commonly used, because the inferences derived from
Jeffreys prior are parameterization-invariant (See Kass \& Wasserman 1996 for detail of
this prior). Under this prior, the conditional distribution of $\lambda$ is
\begin{eqnarray}p_J(\lambda_A=x|n_A=a)&=&\frac{P(n_A=a|\lambda_A=x)(1/\sqrt{x})}{\int_0^{\infty}P(n_A=a|\lambda_A=t)(1/\sqrt{t}){\rm d}t}\nonumber\\
&=&\frac{2^ax^{a-1/2}e^{-x}}{(2a-1)!!\sqrt{\pi}},
\end{eqnarray}
and
\begin{eqnarray}p_J(\lambda_B=x|n_B=b)&=&\frac{2^bx^{b-1/2}e^{-x}}{(2b-1)!!\sqrt{\pi}}.
\end{eqnarray}

To account for the background contamination, suppose that $a$ is the count
corresponding to a Poisson process $A$ with the addition of a Poisson background
process $A_{bkg}$, the expectation value of the process $A_{bkg}$ is assumed to be
known as $\lambda_{Ab}$. According to the properties of Poisson processes, the sum of
two Poisson processes is also a Poisson process with the parameter
$\lambda_A+\lambda_{Ab}$. So the probability of the total count $n$ is
$P(n=a|\lambda_A=x,\lambda_{Ab})=\frac{(x+\lambda_{Ab})^a}{a!}e^{-(x+\lambda_{Ab})}$.
Apply the Bayesian assumption and the uniform prior distribution assumption, we obtain
the conditional distribution of $\lambda_A$, as follows.
\begin{eqnarray}
p_u(\lambda_A=x|n_A=a,\lambda_{Ab})&=&\frac{P(n_A=a|\lambda_A=x,\lambda_{Ab})p_u(\lambda_A=x)}{\int_0^{\infty}P(n_A=a|\lambda_A=t,\lambda_{Ab})p_u(\lambda_A=t){\rm
d}t}\nonumber\\&=&\frac{P(n_A=a|\lambda_A=x,\lambda_{Ab})}{\int_0^{\infty}P(n_A=a|\lambda_A=t,\lambda_{Ab}){\rm
d}t}\nonumber\\&=&\frac{(x+\lambda_{Ab})^ae^{-x}}{a!\sum\limits_{k=0}^a(\frac{\lambda_{Ab}^k}{k!})}.
\end{eqnarray}
When $\lambda_{Ab}$ is much smaller than $a$, this result is same
as equation(2). Also we can obtain the conditional distribution
under the Jeffreys prior distribution assumption,
\begin{eqnarray}
p_J(\lambda_A=x|n_A=a,\lambda_{Ab})
&=&\frac{(x+\lambda_{Ab})^ae^{-x}(1/\sqrt{x})}{\int_0^{\infty}(t+\lambda_{Ab})^ae^{-t}(1/\sqrt{t})dt},
\end{eqnarray}
and it is same as equation(4) when $\lambda_{Ab}$ is much smaller
than $a$.

\bigskip
\section{Estimate the Hardness Ratio}

There are two different definitions of hardness ratio,
$R=\lambda_A/\lambda_B$ and
$HR=(\lambda_B-\lambda_A/)/(\lambda_B+\lambda_A)$. In traditional
method, the estimate of $R$ and $HR$ are $a/b$ and $(b-a)/(b+a)$
respectively, and the errors are propagated under the Gaussian
distribution, i.e.,
\begin{eqnarray}
\sigma_R&=&\frac{a}{b}\sqrt{\frac{\sigma_a^2}{a^2}+\frac{\sigma_b^2}{b^2}},\nonumber\\
\sigma_{HR}&=&\frac{2\sqrt{b^2\sigma_a^2+a^2\sigma_b^2}}{(b+a)^2}.
\end{eqnarray}

Here we propose a method to estimate the hardness ratio based on
the Bayesian method. Both the uniform prior and the Jeffreys prior
will be used. When $\lambda_{Ab}$ is much smaller than $a$, we use
equation(2) (under the uniform prior) or equation(4) (under the
Jeffreys prior) to estimate $R=\lambda_A/\lambda_B$ and
$HR=(\lambda_B-\lambda_A)/(\lambda_B+\lambda_A )$.

First we assume the uniform prior of $\lambda$. For the
conditional distribution function of $\lambda_A/\lambda_B$,
\begin{eqnarray} P_u(\frac{\lambda_A}{\lambda_B}\leq z)&=&
\int_0^\infty
P(\frac{\lambda_A}{\lambda_B}\leq z|\lambda_B=y)p_u(\lambda_B=y){\rm d}y\nonumber\\
&=& \int_0^\infty (\int_0^{zy}\frac{x^ae^{-x}}{a!}{\rm
d}x)\frac{y^be^{-y}}{b!}{\rm d}y.
\end{eqnarray}
For the conditional probability density function of
$\lambda_A/\lambda_B$,
\begin{eqnarray}
p_u(\frac{\lambda_A}{\lambda_B}=z) &=& \frac{\rm d}{{\rm d} z
}P_u(\frac{\lambda_A}{\lambda_B}\leq z) \nonumber\\
&=& \frac{\rm d}{{\rm d} z }\int_0^\infty
(\int_0^{zy}\frac{x^ae^{-x}}{a!}{\rm d}x)\frac{y^be^{-y}}{b!}{\rm
d}y\nonumber\\
&=& \int_0^\infty (\frac{\partial}{\partial z }
\int_0^{zy}\frac{x^ae^{-x}}{a!}{\rm d}x)\frac{y^be^{-y}}{b!}{\rm
d}y\nonumber\\
&=& \int_0^\infty
\frac{{(zy)}^ae^{-zy}}{a!}y\frac{y^be^{-y}}{b!}{\rm d}y\nonumber\\
&=& \frac{z^a(a+b+1)!}{(z+1)^{a+b+2}a!b!}.
\end{eqnarray}
This distribution is shown in Fig. 2 when $a=4$, $b=3$. It is easy
to verify that the distribution is normalized,
\begin{eqnarray}
\int_0^\infty \frac{z^a(a+b+1)!}{(z+1)^{a+b+2}a!b!}{\rm d}z=1.
\end{eqnarray}

For the hardness ratio $HR$, the probability distribution of this
hardness ratio is given by,
\begin{eqnarray}
P_u(\frac{\lambda_B-\lambda_A}{\lambda_B+\lambda_A}\leq z)&=&
\int_0^\infty
P(\frac{\lambda_B-\lambda_A}{\lambda_B+\lambda_A}\leq z|\lambda_A=x)p_u(\lambda_A=x){\rm d}x\nonumber\\
&=& \int_0^\infty
P(\lambda_B\leq \frac{1+z}{1-z}\lambda_A|\lambda_A=x)p_u(\lambda_A=x){\rm d}x\nonumber\\
&=& \int_0^\infty
(\int_0^{\frac{1+z}{1-z}x}\frac{y^be^{-y}}{b!}{\rm
d}y)\frac{x^ae^{-x}}{a!}{\rm d}x.
\end{eqnarray}
The conditional probability density function is given by,
\begin{eqnarray}
p_u(\frac{\lambda_B-\lambda_A}{\lambda_B+\lambda_A}=z) &=&
\frac{\rm d}{{\rm d} z
}P_u(\frac{\lambda_B-\lambda_A}{\lambda_B+\lambda_A}\leq z) \nonumber\\
&=& \frac{\rm d}{{\rm d} z }\int_0^\infty
(\int_0^{\frac{1+z}{1-z}x}\frac{y^be^{-y}}{b!}{\rm
d}y)\frac{x^ae^{-x}}{a!}{\rm d}x\nonumber\\
&=& \int_0^\infty (\frac{\partial}{\partial z
}\int_0^{\frac{1+z}{1-z}x}\frac{y^be^{-y}}{b!}{\rm
d}y)\frac{x^ae^{-x}}{a!}{\rm d}x\nonumber\\
&=& \int_0^\infty
\frac{{(\frac{1+z}{1-z}x)}^be^{-\frac{1+z}{1-z}x}}{b!}\frac{2x}{(1-z)^2}\frac{x^ae^{-x}}{a!}{\rm d}x\nonumber\\
&=& \frac{(1-z)^a(1+z)^b(a+b+1)!}{2^{(a+b+1)}a!b!}.
\end{eqnarray}
This distribution is shown in Fig. 3 when $a=4$, $b=3$. It is easy
to verify that the distribution is normalized,
\begin{eqnarray}
\int_0^\infty \frac{(1-z)^a(1+z)^b(a+b+1)!}{2^{(a+b+1)}a!b!}{\rm
d}z=1.
\end{eqnarray}

\begin{figure}[tb]
 \begin{minipage}[t]{0.48\linewidth}
  \centering
  \includegraphics[scale=0.41]{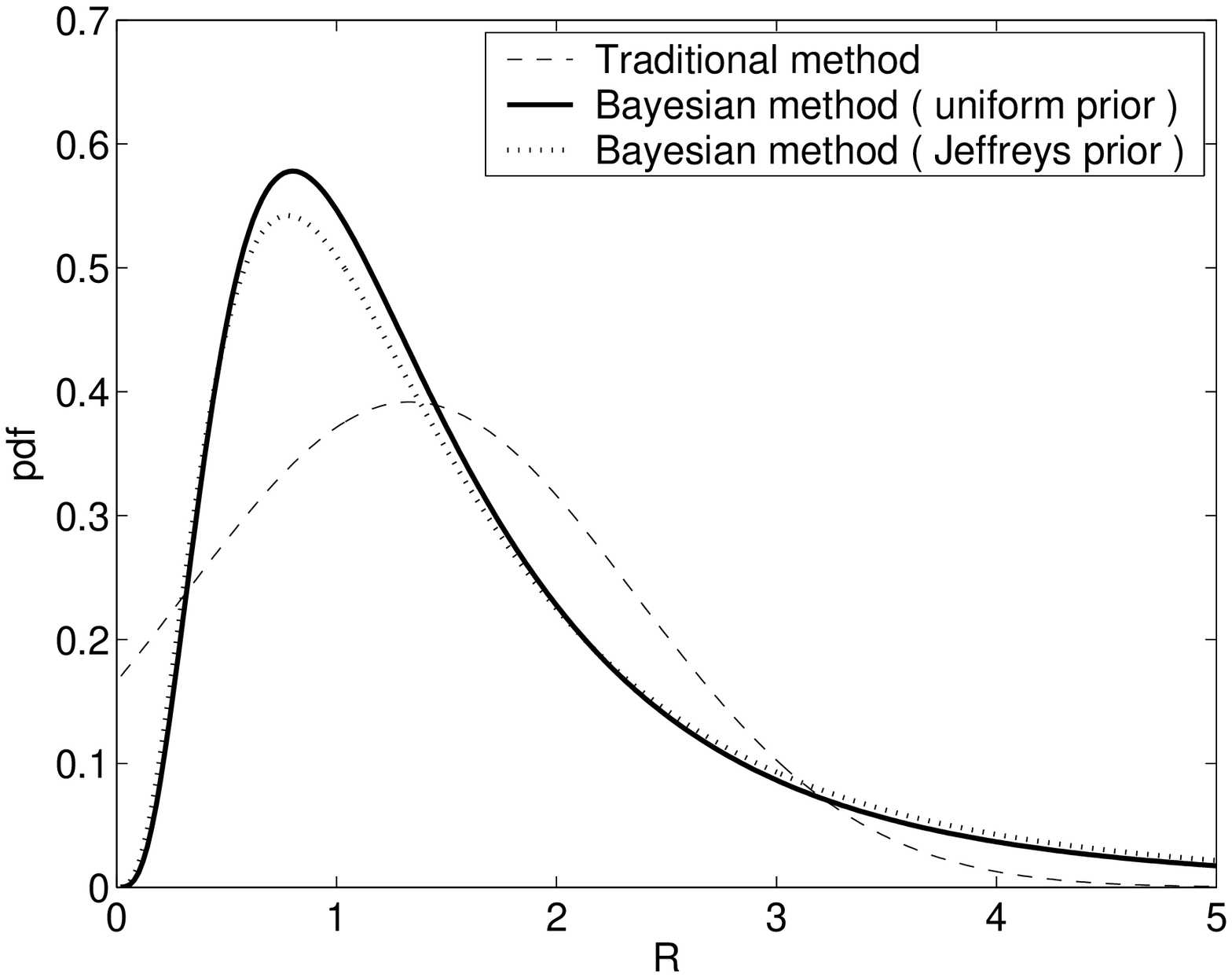}
  \vspace{-0.2in}
  \caption[]{{The probability density function of
$\lambda_A/\lambda_B$ when $a=4$, $b=3$. The solid line represents
the distribution derived by our proposed method under the uniform
prior, the doted line represents the distribution derived by our
proposed method under the Jeffreys prior, the dashed line
represents the Gaussian distribution derived by the traditional
method.}}
  \label{fig:side:a}
 \end{minipage}\ \ \ \ \ \ \
 \begin{minipage}[t]{0.48\linewidth}
  \centering
  \includegraphics[scale=0.41]{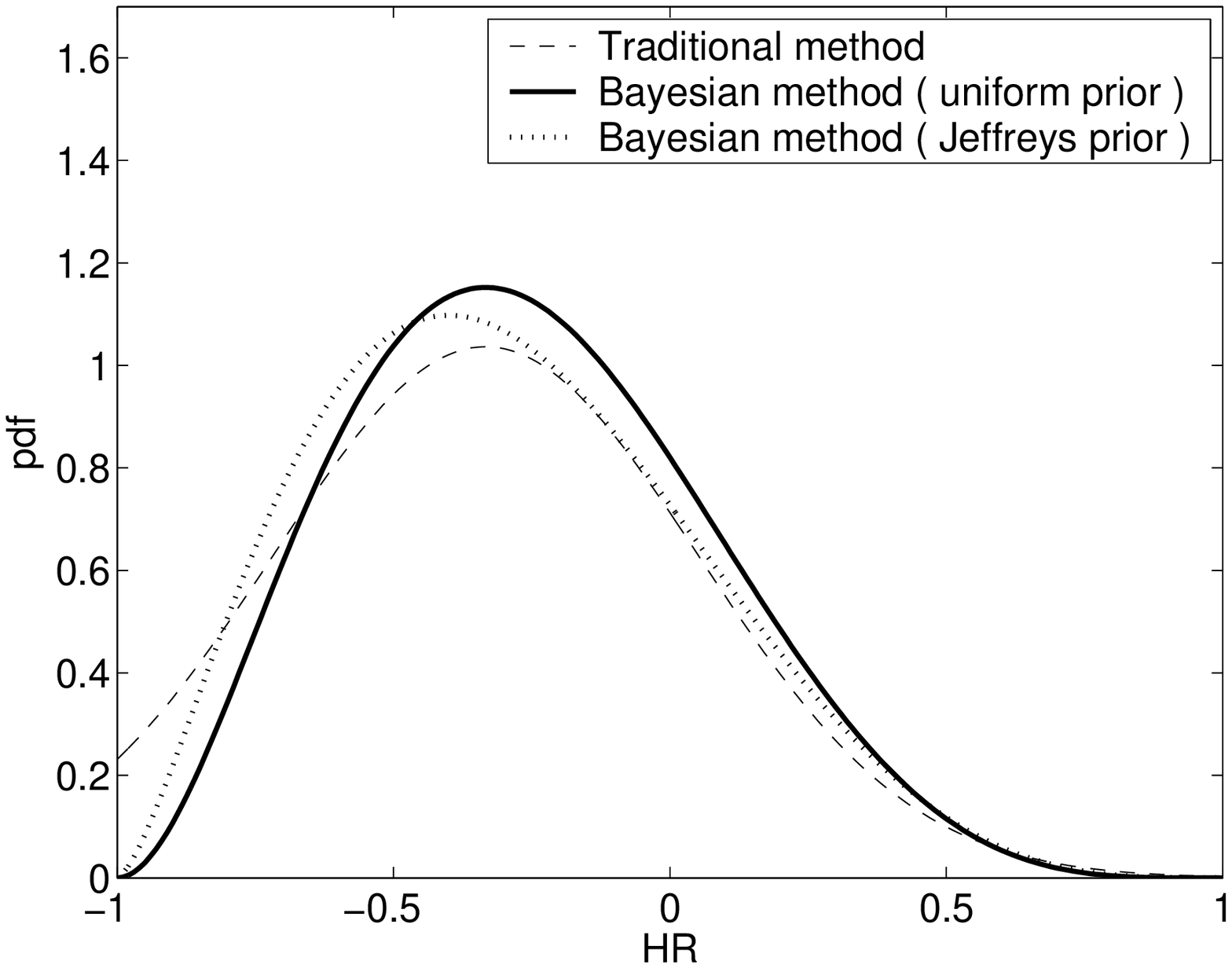}
  \vspace{-0.2in}
  \caption[]{{The probability density function of $(\lambda_B-\lambda_A)/(\lambda_B+\lambda_A)$ when
  $a=4$, $b=3$.The solid line represents
the distribution derived by our proposed method under the uniform
prior, the doted line represents the distribution derived by our
proposed method under the Jeffreys prior, the dashed line
represents the Gaussian distribution derived by the traditional
method.}}
  \label{fig:side:b}
 \end{minipage}
\end{figure}

Similarly, we obtain the conditional probability density function
of $R$ and $HR$ under the Jeffreys prior assumption as follows,
\begin{eqnarray}
p_J(R=z)&=&\frac{z^{a-1/2}2^{a+b}(a+b)!}{(z+1)^{a+b+1}(2a-1)!!(2b-1)!!\pi},
\end{eqnarray}
and
\begin{eqnarray}
p_J(HR=z)&=&\frac{(1-z)^{a-1/2}(1+z)^{b-1/2}(a+b)!}{(2a-1)!!(2b-1)!!\pi}.
\end{eqnarray}

Using the result of $p(R=z)$ and $p(HR=z)$, we can estimate $R$
and $HR$.

Under the uniform prior assumption, since we have only one
observation, we take the most probable value as the estimate of
$R=\lambda_A/\lambda_B$, denoted as $(z_R)_u$. Let
\begin{eqnarray} \frac{\partial}{\partial
z}p_u(\frac{\lambda_A}{\lambda_B}=z|n_A=a, n_B=b)&=&0,
\end{eqnarray}
we obtain,
\begin{eqnarray}{(z_R)_u}&=&\frac{a}{b+2}.
\end{eqnarray}
Similarly, we obtain the most probable value as the estimate of
HR:
\begin{eqnarray}{(z_{HR})_u}&=&\frac{b-a}{b+a}.
\end{eqnarray}

Similarly, we get the most probable value of $R$ and $HR$ under
the Jeffreys prior assumption,
\begin{eqnarray}{(z_R)_J}&=&\frac{a-1/2}{b+3/2},
\end{eqnarray}
and
\begin{eqnarray}{(z_{HR})_J}&=&\frac{b-a}{b+a-1}.
\end{eqnarray}

The highest posterior density (HPD) interval is used to give the
error bars. The HPD interval under the confidence level $\alpha$
is the range of values which contain a fraction $\alpha$ of the
probability, and the probability density within this interval is
always higher than that outside the interval.

There are other point estimates and error estimates. For example,
the mean value and the equal tailed interval. Since the
distributions of $R$ and $HR$ are obtained, these alternative
estimates can be easily derived.

When $\lambda_{Ab}$ cannot be ignored, equation (6) or equation
(7) can be used to estimate $R$ and $HR$. In this situation, it is
difficult to give a simple analytic distribution function like
equation (10) or equation (13); in this case, one can only use
numerical integration to obtain the distribution of $R$ and $HR$,
and then do the estimate.

\section{Frequency Properties of Intervals}

We use the Monte Carlo simulations to investigate the statistical
properties of our result, and compare them with the traditional
methods.

First we set $\lambda_A$ and $\lambda_B$. Do Poisson sampling for
$N$ times and each time we get $a$ and $b$ respectively. Each time
we estimate $R$ and $HR$ using two kinds of methods. Finally we
obtain that, for two methods, the mean square error of the point
estimate, the coverage rate (the percentage of times during which
the confidence interval contains the real value), and the mean
confidence interval.

The simulations contain two cases: low counts and high counts. In
case 1, we first set $\lambda_A=3$ and $\lambda_B=3$, then set
$\lambda_A=4$ and $\lambda_B=3$. In case 2, we first set
$\lambda_A=20$ and $\lambda_B=20$, then set $\lambda_A=20$ and
$\lambda_B=15$. The confidence level in the simulations is 90\%.
The results of the simulations are shown in table 1.

\clearpage
\begin{table}[t]
  \begin{center}
    \caption{Statistical Properties of Our Method and Traditional Method.}
  \medskip
  \begin{tabular}{ccc|cc|cc} \hline\hline
  & & & \mdivi{Low counts} & \multi{High counts}\\
  \cline{4-7}
  & & & {1:1} & {4:3} & {1:1} & {4:3}\\
  \hline
  \divi{}&{our method} & R & 0.5291 & 0.6872 & 0.3066 & 0.4397\\
  \cline{2-7}
  \divi{\upa{mean square error}}&{traditional method} & R & 1.2399 & 1.5308 & 0.3628 & 0.5767\\
  \hline
  \divi{}&{\upd{our method}} & R  & 94.62\% & 89.13\% & 93.00\% & 90.00\% \\
  \cline{3-7}
  \divi{}&{\upc{(uniform prior)}}    & HR & 94.38\%  & 91.49\% & 85.90\% & 85.00\% \\
  \cline{2-7}
  \divi{}&{\upd{our method}} & R  & 92.27\% & 92.35\% & 91.30\% & 87.87\% \\
  \cline{3-7}
  \divi{}&{\upc{(Jeffreys prior)}}    & HR & 82.82\%  & 87.52\% & 87.50\% & 84.62\% \\
  \cline{2-7}
  \divi{\up{coverage rate}}&{\upd{traditional}} & R  & 84.70\% & 84.82\%  & 90.29\% & 89.46\%  \\
  \cline{3-7}
  \divi{}&{\upc{method}}   & HR & 88.22\% & 85.34\%  & 89.10\% & 88.26\%  \\
  \hline
  \divi{}&{\upd{our method}} & R  &  3.40 &  4.67    & 1.20  & 1.60  \\
  \cline{3-7}
  \divi{}&{\upc{(uniform prior)}}    & HR & 1.06  & 1.00   & 0.50 & 0.53 \\
  \cline{2-7}
  \divi{}&{\upd{our method}} & R  &  4.05 &  5.56    & 1.17  & 1.80  \\
  \cline{3-7}
  \divi{\upb{mean }}&{\upc{(Jeffreys prior)}}    & HR & 1.13  & 1.05   & 0.50 & 0.56 \\
  \cline{2-7}
  \divi{\upb{confidence interval}}&{\upd{traditional}} & R  & 4.91 & 5.90  & 1.17 & 1.78  \\
  \cline{3-7}
  \divi{}&{\upc{method}}   & HR & 1.43 & 1.23  & 0.52 & 0.55  \\
  \hline
  \end{tabular}
  \label{park:tbl:coverage}
  \end{center}
\end{table}

From the simulation results, we notice that our proposed method is
more reliable than the traditional method when the counts are low.
The reason is that the traditional method is based on using the
Gaussian distribution to approach the Poisson distribution, which
is not reliable when the counts are low.

\section{Application to $N_H$ Estimation}

\begin{figure}[tb]
\centerline{\epsfxsize 80mm \plotone{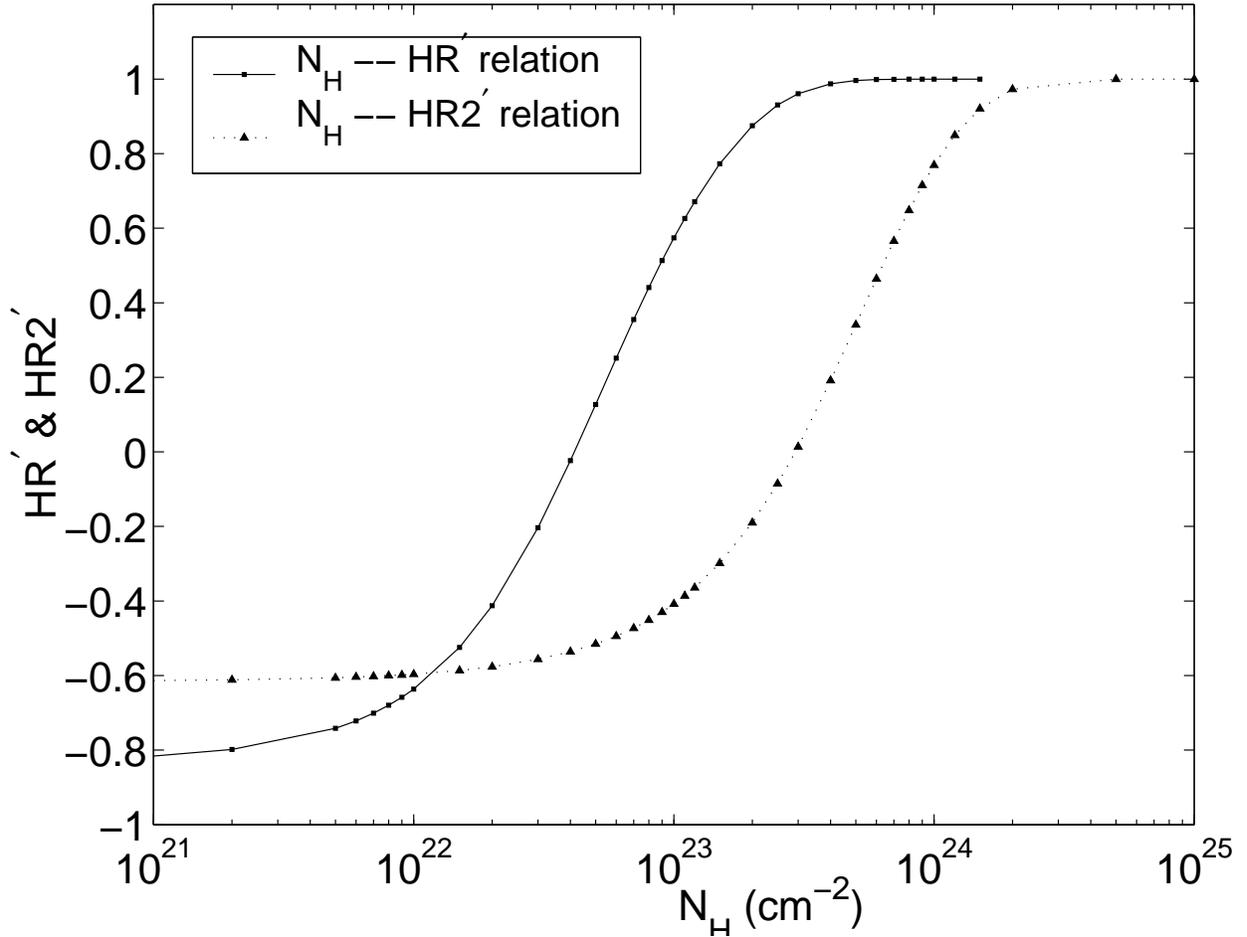}} \caption[]{Relation
between $N_H$ and $HR'$ \& $HR2'$. The solid line with squares is
the relation of $N_H$ and $HR'$, while the dotted line with
triangles is the one of $N_H$ and $HR2'$ (Wu et al. 2006)}
\label{fig7}
\end{figure}

Here we propose a method to estimate the hydrogen column density
using data obtained with the {\it Chandra} X-ray observatory.
Because of the high angular resolution of {\it Chandra}, a
positive detection of a point source during a survey observation
only requires several counts. Therefore our method will be more
reliable when estimating the hardness ratio than the traditional
method. The detail procedure of this application can be found in
another paper (Wu et al. 2006). The basic idea is introduced as
follows.

The basic procedure consists of the following three steps: (1)
calculate the relationship between the hardness ratios and $N_H$
values under certain spectral model; (2) estimate hardness ratios
according to observed counts in different wavebands; and (3)
interpolate the $N_H$ values and error intervals from hardness
ratios.

According to the most likely physical nature of the sources, we
can assume a spectral model (e.g. power law with photon index
$\Gamma=2$ for typical X-ray binaries). Then we can use {\it
PIMMS} (http://cxc.harvard.edu/toolkit/pimms.jsp) tool to
calculate the relationship between hydrogen column density and the
hardness ratio. Using {\it PIMMS}, one can get the count rate in
certain energy band under a given X-ray spectrum and hydrogen
column density. The count rate in the given energy band is just
the $\lambda$ parameter in this energy band; this was our original
movivation of defining the hardness ratios in terms of the
$\lambda$ parameter. The calculated $N_H$ (hydrogen column
density) --- $HR$ relationships are shown in Fig. 7 for three
different energy bands of $A$ (1 - 3 keV), $B$ (3 - 5 keV) and $C$
(5 - 8 keV) for a {\it Chandra} ACIS-I observation (Wu et al.
2006) respectively:
$HR'=\frac{\lambda_B-\lambda_A}{\lambda_B+\lambda_A}$,
$HR2'=\frac{\lambda_C-\lambda_B}{\lambda_C+\lambda_B}$. From Fig.
7, we can see that $HR'$ or $HR2'$ is more appropriate for $N_H <
2\times 10^{23}{\rm cm}^{-2}$ or $N_H
> 2\times 10^{23}{\rm cm}^{-2}$, respectively. Having the value and error
interval of $HR$, we can finally do linear interpolation on curves
in Fig. 7 to obtain the value and error interval of hydrogen
column density.

\section{Summary \& Discussion}

First we give the conditional probability distribution of
$\lambda$ parameter under certain counts in a Poisson process
using Bayesian statistics. According to this result we derive the
probability density function of two kinds of hardness ratios. We
take the most probable values as the estimate of hardness ratios
and the HPD intervals as the error intervals. Then we use Monte
Carlo simulations to investigate the statistical properties of our
results, and find that our method is more reliable than the
traditional method when the counts are low. Finally we show how to
estimate the hydrogen column density using hardness ratios.

Our method developed in this paper provides a way to estimate the
hydrogen column density of sources which are too faint to do
spectral fitting. However the spectral shape for these sources
must be assumed {\it a prior}. This method is especially
convenient for a sample of faint sources with similar spectra.

After this paper has been submitted initially on 06-03-29, we
noticed another submitted paper (Park et al. 2006) which discusses
the same statistical problem as we have done in this paper. In
that paper the authors also used the Bayesian method to estimate
the hardness ratio, and showed some applications on quiescent
Low-Mass X-ray Binaries, the evolution of a flare, etc, therefore
justifying the wide range of applications of such a statistical
problem. Since the strict analytic solution of the hardness ratio
distribution does not exist for general situations, the authors
suggested methods by Monte Carlo and numerical integration to
obtain the distribution in that paper. In our paper we find simple
analytic solutions of the probability density functions of
hardness ratios for the situations in which the background can be
ignored. This will be useful and convenient for some applications,
such as {\it Chandra} data in which background can be ignored for
hardness ratio estimation of point sources.

Finally, we note, under the advise of the referee, that in 1980s
some studies have been done on the ratio of Poisson means both
from a frequentist standpoint (James \& Roos 1980) and from a
Bayesian standpoint (Helene 1984 and Prosper 1985). In this paper
we used the Bayesian method under the uniform prior and the
Jeffreys prior, made extensive comparisons between this method and
traditional method, aiming explicitly at applications in
astrophysics.

\begin{acknowledgements}
Xie Chen read the first draft carefully and gave many helpful suggestions, especially
on English writing. We are particularly grateful to the referee for his pointing out
relevant historical literature in other fields and suggesting using Jeffreys prior.
This study is supported in part by the Special Funds for Major State Basic Research
Projects and by the National Natural Science Foundation of China (project no.10233030,
10327301 and 10521001).
\end{acknowledgements}

\end{document}